\documentstyle[prl,aps,amssymb,epsfig]{revtex}

\begin{document}

\newcommand{\lsco}{La$_{2-x}$Sr$_{x}$CuO$_{4}$}
\newcommand{\cuo}{CuO$_{2}$}
\newcommand{\lscoUD}{La$_{1.88}$Sr$_{0.12}$CuO$_{4}$}
\newcommand{\lscoOD}{La$_{1.85}$Sr$_{0.15}$CuO$_{4}$}
\newcommand{\ybco}{YBa$_{2}$Cu$_{3}$O$_{7-\delta}$}
\newcommand{\ba}{{\bf a}}
\newcommand{\bc}{{\bf c}}
\newcommand{\bac}{{\bf ac}}
\newcommand{\Tc}{$T_{c}$}
\newcommand{\somega}{$\sigma(\omega)$}
\newcommand{\iOhmcm}{$\Omega^{-1}$cm$^{-1}$}
\newcommand{\iOhmcmcm}{$\Omega^{-1}$cm$^{-2}$}
\newcommand{\mkm}{$\mu$m}
\newcommand{\degree}{$^{\circ}$}
\newcommand{\icm}{cm$^{-1}$}
\newcommand{\etal}{{\it et al.}}

 \draft

\twocolumn[\hsize\textwidth\columnwidth\hsize\csname@twocolumnfalse\endcsname


\title{The \bc-axis optical sum rule and possible new collective mode in \lsco}

\author{A.~B.~Kuzmenko$^{1}$, N.~Tombros$^{1}$, H.~J.~A.~Molegraaf$^{1}$, M.Gr\"uninger$^{2}$,
D.~van~der~Marel$^{1}$ and S.~Uchida$^{3}$}

\address{$^{1}$Material Science Center, University of Groningen, Nijenborgh 4, 9747AG, Groningen, The Netherlands}
\address{$^{2}$II Physical Institute, University of Cologne, 50937 Cologne, Germany}
\address{$^{3}$Department of Superconductivity, University of Tokyo, Bunkyo-ku, Tokyo 113, Japan}

\maketitle

\begin{abstract}

We present the \bc-axis optical conductivity $\sigma_{1c}(\omega,
T)$ of underdoped ($x$ = 0.12) and optimally doped ($x$ = 0.15)
\lsco\ from 4 meV to 1.8 eV obtained by a combination of
reflectivity and transmission spectra. In addition to the opening
of the superconducting gap, we observe an {\it increase} of
conductivity above the gap up to 270 meV with a maximal effect at
about 120 meV. This may indicate a new collective mode at a
surprisingly large energy scale. The Ferrell-Glover-Tinkham sum
rule is violated for both doping levels. Although the relative
value of the violation is much larger for the underdoped sample,
the absolute increase of the low-frequency spectral weight,
including that of the condensate, is higher in the optimally doped
regime. Our results resemble in many respects the observations in
\ybco.

\end{abstract}
]

\vskip2pc

%

The charge transport between the \cuo\ planes in the high-\Tc\
cuprates forms one of the most intriguing puzzles of these
materials. On the one hand the essential common physics seems to
lie in the collective behavior of holes doped into a 2D Mott
insulator, while the interplane conductivity strongly depends on
the interlayer chemistry, which varies dramatically among
different members of the high-\Tc\ family. On the other hand, it
is the in-plane 'confinement' that preserves a possibility of
significant lowering of the \bc-axis kinetic energy (KE) as the 3D
coherent movement of the Cooper pairs is restored below
\Tc\cite{Anderson97,Munzar01}, which, however, was shown to be a
small effect in some single-layer compounds
\cite{Schuetzmann97,Moler98}. The models based on the KE lowering
in the superconducting (SC) state predict the violation of the
Ferrell-Glover-Tinkham (FGT) optical sum rule\cite{Tinkham59},
which means that the spectral weight (SW) of the SC condensate is
collected not only from energies of the order of $2\Delta$ but
also from higher frequencies. For the \bc-axis the violation has
been experimentally tested in \ybco\ (YBCO) over a broad doping
range and a significant increase of the relative violation value
was found in the underdoped regime \cite{Basov01}. In the case of
\lsco\ (LSCO) a value of 50\% has been reported for a slightly
underdoped sample \cite{Basov99}, although the doping dependence
is still not known. Thus, the \bc-axis KE lowers in the SC state,
which, however, leaves room for debate whether this is a
by-product or the very reason for superconductivity
\cite{Anderson97,Munzar01,Hirsch00,Ioffe99}.

The conclusiveness of the sum rule analysis greatly depends on a
reliable value of $\sigma_{1}(\omega,T)$. However, the usually
employed Kramers-Kronig (KK) analysis of reflectivity seriously
lacks accuracy due to the insensitivity of reflection to small
absorption details and an ambiguity of the data extrapolation.
This is a major problem for LSCO with a relatively small \bc-axis
electronic conductivity \cite{Uchida96}. In this Letter we report
on our optical study of underdoped (UD) and optimally doped (OpD)
LSCO by a combination of reflection and transmission spectroscopy
which overcomes many limitations of the KK analysis. We measured
the transmission (${\bf E}\parallel{\bf c}$) through free-standing
thin crystals to avoid complications due to a substrate. Here we
focus on the low-temperature data well below the HTT-LTO phase
transition.

\begin{figure}
    \centerline{\psfig{figure=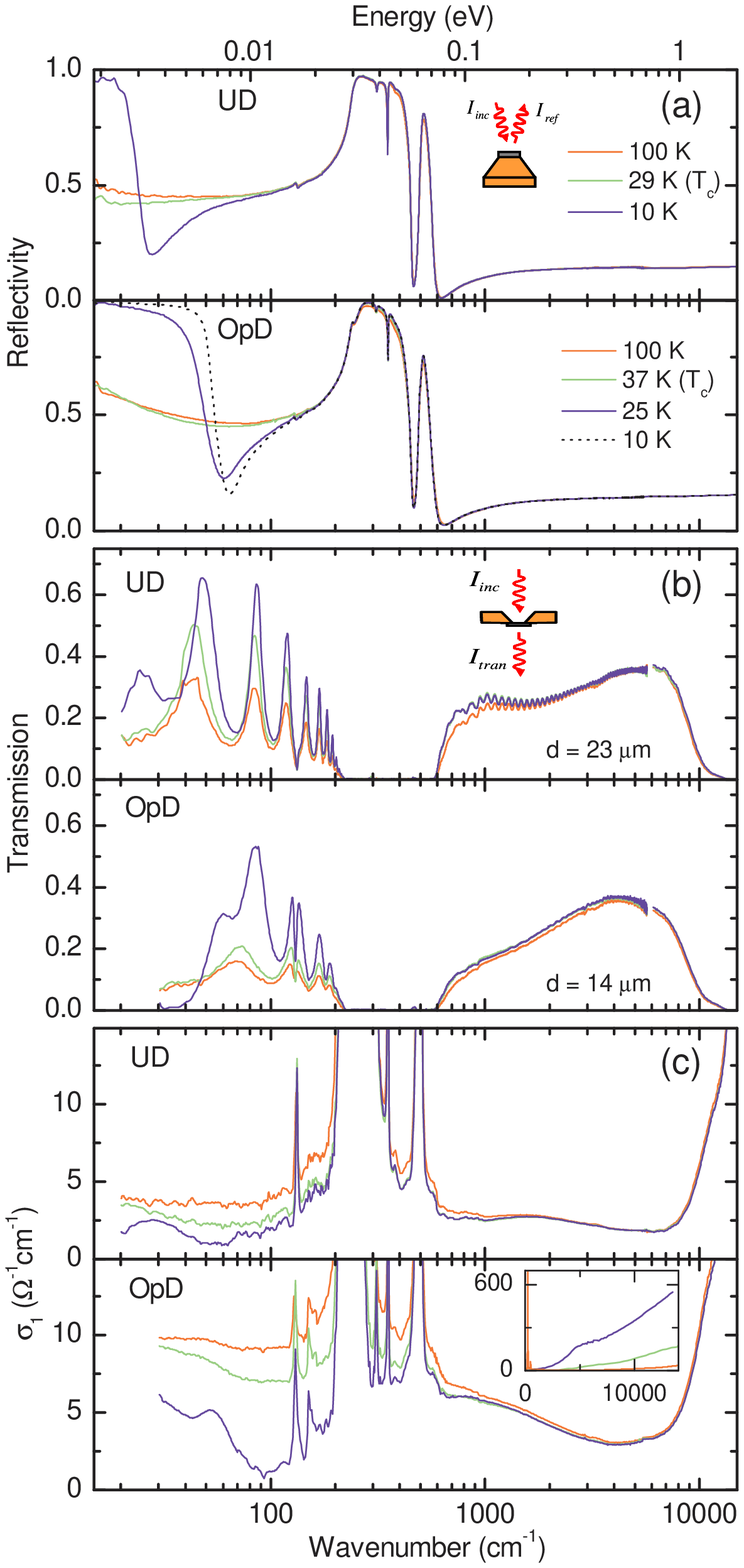,width=8cm,clip=}}
    \caption{The \bc-axis reflectivity (a), transmission (b) and optical conductivity
    (c) of the UD ($x$=0.12) and the OpD ($x$=0.15) LSCO samples. The same colors at the same doping
    correspond to the same temperature. The inset compares $\sigma_{1}(\omega,300K)$ (red) with the previous
    results based on the KK analysis (Ref.\protect\cite{Basov95} - blue, Ref.\protect\cite{Tamasaku94} - green).}
     \label{Fig1}
\end{figure}

The \lsco\ single crystals ($x$=0.12, \Tc=29 K and $x$=0.15,
\Tc=37 K) were grown by the same method as described in Ref.
\cite{Nakamura93}. The \bc-axis was within 1\degree\ in the {\bf
ac} sample surface, as determined by Laue diffraction. Initially
the reflectivity ($R$) spectra (15 - 6000 \icm) have been obtained
using an FT-IR spectrometer on thick samples with the surface area
of 20-25 mm$^{2}$ mounted on a copper cone (see
Fig.\ref{Fig1}(a)). The {\it in-situ} gold evaporation was used as
a reference. The spectra match very well with the previous data
\cite{Uchida96}, which were used to continue curves to higher
frequencies. Then each sample was attached to a supporting plane
and reduced to a thin platelet by polishing with diamond paper
with a roughness down to 0.1 \mkm. The platelets were made 23
\mkm\ and 14 \mkm\ thick with areas of 1.3 and 1.0 mm$^{2}$ for
$x$ = 0.12 and 0.15 correspondingly. The thickness was chosen in
order to optimize the transmission sensitivity on the base of an
anticipated value of $\sigma_{1}$. Finally the samples were
unglued from the support and mounted on a copper mask. The
\bc-axis transmission ($T$) spectra (see Fig.\ref{Fig1}(b)) have
been measured using an FT-IR spectrometer (20 - 6000 \icm) and a
grating-type spectrometer (6000 - 15000 \icm). At higher
frequencies the samples are not transparent presumably due to the
charge-transfer absorption, at lower energies the diffraction
effects mask the data. The transmission for
$\bf{E}\parallel\bf{a}$ was unmeasurably small, indicating the
absence of any 'pinhole' leakage. All the measurements were done
in a home-made optical cryostat with a temperature-stable sample
position. The absolute systematic error bars of $R$ and $T$ are
about 0.01 and 0.03 respectively. The accuracy of their relative
temperature difference, is at least one order of magnitude better.
The temperature precision is about 1 K \cite{TempCal}.

The measured $R(\omega)$ and $T(\omega)$ are functions of
$\epsilon(\omega)=\epsilon_{1}(\omega)+i\epsilon_{2}(\omega)$:
\begin{equation}
R=|r|^2, T=\left|\frac{(1-r^{2})t}{1-r^{2}t^{2}}\right|^2,
r=\frac{1-\sqrt{\epsilon}}{1+\sqrt{\epsilon}},
t=e^{i\frac{\omega}{c}\sqrt{\epsilon}d},
\end{equation}
\noindent where $d$ is the sample thickness. These relations can
be applied to derive $\epsilon_{1}$ and $\epsilon_{2}$ from $R$
and $T$ for frequencies above 650 \icm\ and below 200 \icm\
without the use of the KK transformation. In the phonon range,
where $T$ is vanishingly small, the KK method still should be
exploited. Here we use a more accurate analysis (which we call
'RT+KK'), which is to find an $\epsilon(\omega)$ which satisfies
the KK relations and delivers the best fit to the experimental
$R(\omega)$ and $T(\omega)$ {\it simultaneously} in the whole
spectral range. It is done by modelling $\epsilon(\omega)$ with a
very large number of oscillators (approaching the number of
experimental points), including the zero-frequency lossless mode
of the SC condensate. The sample thickness is one of the variable
parameters which is determined by fitting of the Fabry-Perot
fringes \cite{Spread}. The two methods give close results. The
optical conductivity $\sigma_{1}(\omega) =
\omega\epsilon_{2}(\omega)/4\pi$ is shown in Fig.\ref{Fig1}(c).
Notably, above 1000 \icm\ it is significantly smaller than the
previously published values based solely on the KK analysis
\cite{Basov95,Tamasaku94} (see inset in Fig.\ref{Fig1}(c)).

\begin{figure}
    \centerline{\psfig{figure=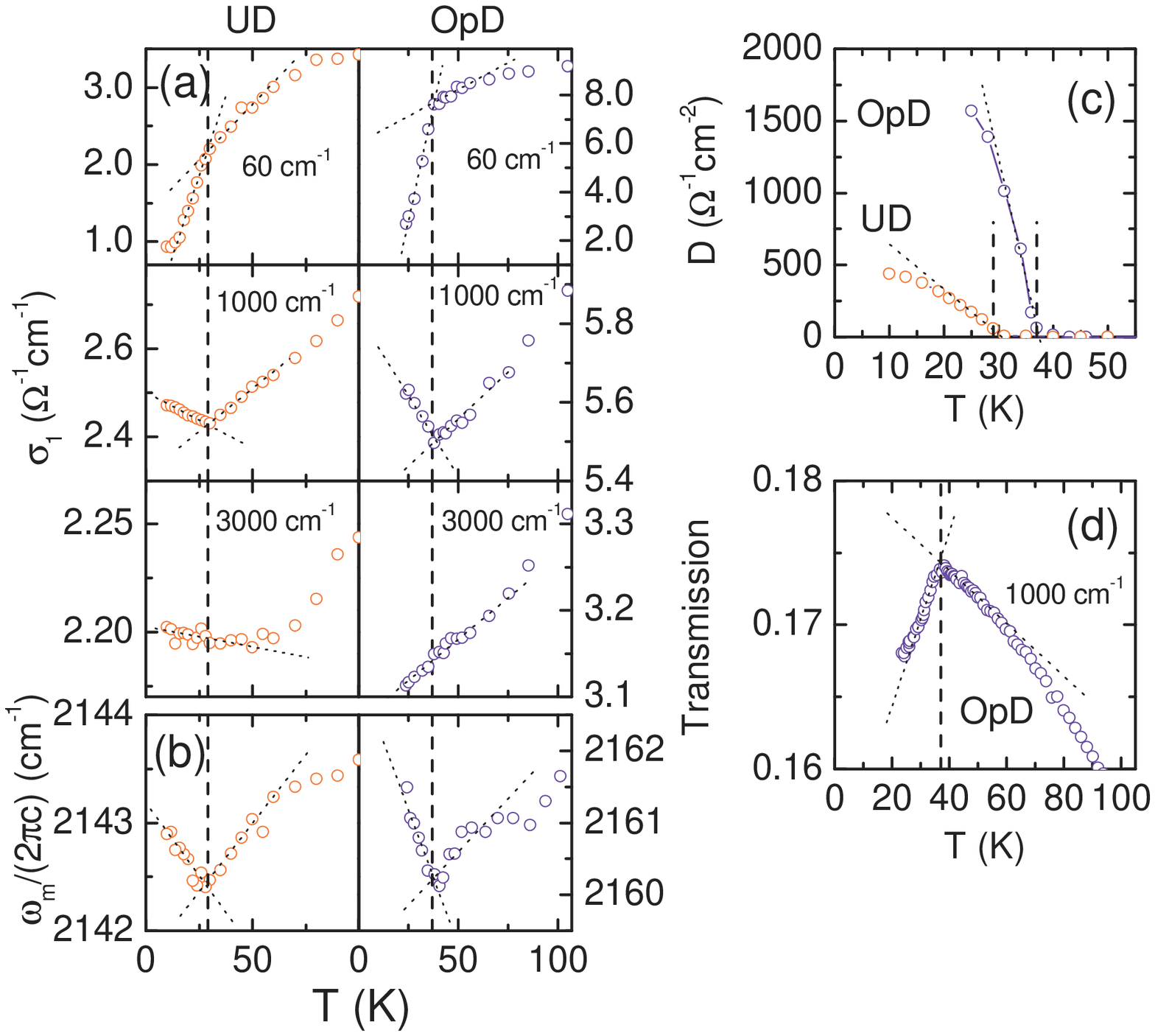,width=8cm,clip=}}
    \caption{The temperature dependence of (a) conductivity at selected
    energies, (b) a Fabry-Perot maximum position at $\approx$ 2150 \icm,
    (c) the condensate SW and (d) the transmission of the OpD sample
    at 1000 \protect\icm\ with a fine temperature resolution.
    The dotted lines denote the curve slopes above and below \Tc;
    the latter is marked by the vertical line.}
     \label{Fig2}
\end{figure}

The conductivity for selected frequencies is shown in
Fig.\ref{Fig2}(a). It is temperature-dependent both in the normal
and the SC state and shows a distinct kink right at \Tc. Because
$\sigma_1(\omega,T)$ has a strong temperature dependence above
$T_c$, the difference
$\Delta\sigma_{1}=\sigma_{1}(T=T_{c})-\sigma_{1}(T\ll T_{c})$ does
not represent purely a superconductivity induced change of
$\sigma_{1}(\omega)$. Instead, we use the {\it slope jump} at \Tc
\begin{equation}
\Delta_{s} \sigma_{1} \equiv \left.\frac{\partial
\sigma_{1}}{\partial T} \right| _{T_{c}+}-\left.\frac{\partial
\sigma_{1}}{\partial T} \right| _{T_{c}-}
\end{equation}
\noindent to characterize the effect of the SC transition. One can
see that the kink is negative at low frequencies, indicating the
opening of the SC gap. However, it becomes {\it positive} at
somewhat higher energies ($\sim$ 1000 \icm, or $\approx$ 120 meV)
and disappears within the noise level above $\Omega_{c}\approx$
2200 \icm\ ($\approx$ 270 meV). The 'anomalous' kink at
intermediate energies is nicely seen in the original transmission
data, which we double-checked with an enhanced temperature
resolution (see Fig.\ref{Fig2}(d)).

\begin{figure}
    \centerline{\psfig{figure=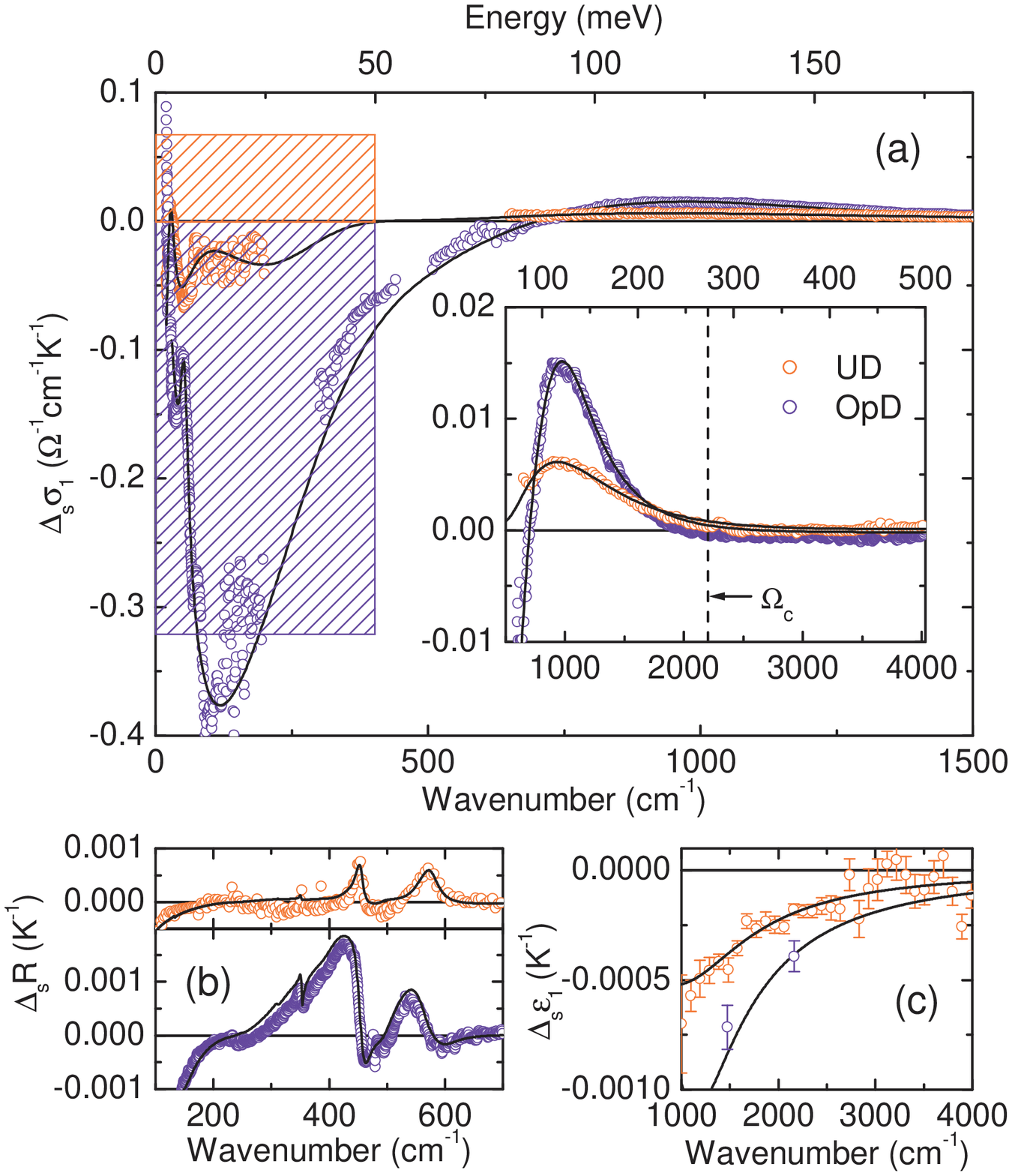,width=8cm,clip=}}
    \caption{(a) $\Delta_{s}\sigma_{1}(\omega)$ (the hatched areas correspond to
    the condensate SW, $\Delta_{s}D$); the inset expands
    the higher frequency region, where transmission allows a very accurate
    determination of $\sigma_{1}(\omega)$. (b) $\Delta_{s}R(\omega)$ in the phonon range.
    (c) $\Delta_{s}\epsilon_{1}(\omega)$ at high frequencies, as derived from the shift of Fabry-Perot fringes.
    The red and blue colors relate to $x$ = 0.12 and 0.15 respectively. The solid
    curves represent the fit as described in the text.}
     \label{Fig3}
\end{figure}

The spectral dependence of the conductivity kink is shown in
Fig.\ref{Fig3}(a). The most accurate value of
$\Delta_{s}\sigma_{1}$ is obtained in regions of non-zero
transmission. The error bars increase in the phonon range,
especially near the very intense peak at 240 \icm, which
complicates the sum rule analysis. The difficulty can be
circumvented by assuming a smooth shape of
$\Delta_{s}\sigma_{1}(\omega)$ and fitting it with a function,
which can be used to 'bridge' the problematic frequency range. The
model function
$\Delta_{s}\epsilon=\Delta_{s}\epsilon_{1}+i\Delta_{s}\epsilon_{2}$
has to satisfy the KK relations. In addition to
$\Delta_{s}\sigma_{1}(\omega)=\Delta_{s}\epsilon_{2}(\omega)\omega/4\pi$
below and above the phonons (Fig. \ref{Fig3}(a)) we fit the
experimentally observed spectrum of $\Delta_{s}R(\omega)$ inside
the phonon region (Fig. \ref{Fig3}(b)) using the relation
\begin{equation}
\Delta_{s}R=(\partial R/\partial
\epsilon_{1})\Delta_{s}\epsilon_{1}+(\partial R/\partial
\epsilon_{2})\Delta_{s}\epsilon_{2},
\end{equation}

\noindent where the derivatives $\partial R/\partial
\epsilon_{1,2}$ are obtained by the RT+KK method at \Tc\ (the idea
is similar to that of the temperature modulation technique).
Important extra information on $\Delta_{s} \sigma_{2}$ (or
$\Delta_{s} \epsilon_{1}$) is given by the Fabry-Perot extrema
positions $\omega_{m}$. The product
$\omega_{m}\sqrt{\epsilon_{1}}d$ is a constant (if $\epsilon_{2}
\ll \epsilon_{1}$), therefore
\begin{equation}
\Delta_{s}\epsilon_{1} = -2\epsilon_{1}\left(\Delta_{s}
\omega_{m}/\omega_{m}+\Delta_{s} d/d\right).
\end{equation}
\noindent The \bc-axis lattice constant does not show a kink at
\Tc \cite{Arai98} so that the spectral dependence of
$\Delta_{s}\epsilon_{1}$ can be directly determined from
$\omega_{m}$'s and fitted together with $\Delta_{s}\sigma_{1}$ and
$\Delta_{s}R$. The high-frequency $\Delta_{s}\epsilon_{1}$ is
especially useful because it is related to the total low-frequency
SW \cite{FormulaDeltaEps}
\begin{equation}
\Delta_{s}\epsilon_{1}(\omega)\approx-8\Delta_{s}\left[A(\Omega_{c})+D\right]\omega^{-2}
\label{DeltaEps}
\end{equation}

\noindent (we adopt notations
$A(\omega)=\int_{0+}^{\omega}\sigma_{1}(\omega^{\prime})d\omega^{\prime}$
for the finite-frequency integrated SW and $D$ for the condensate
SW). As the Fabry-Perot pattern is less damped for $x$=0.12 due to
the smaller thickness spread, the error bars of
$\Delta_{s}\epsilon_{1}$ are much smaller in this case. We
obtained a reasonably good fit of $\Delta_{s}\sigma_{1}$,
$\Delta_{s}R$ and $\Delta_{s}\epsilon_{1}$ simultaneously (see
Fig.\ref{Fig3}), which is prerequisite for a reliable sum rule
examination.

The superconducti\-vity-driven change of $\sigma_{1}$ below \Tc\
(detectable for $\omega<\Omega_{c}$) is governed by two factors:
(i) the SC gap opening and (ii) the increase of SW above the gap.
The former is much more pronounced for the OpD sample ($\sim$ 8
times larger effect). It is therefore remarkable that the latter
is only 1.8-2 times smaller for the UD sample. In principle, a
slight conductivity increase above the gap is expected in BCS
theory in the clean limit. However, the incoherent \bc-axis
transport is far in the dirty limit, especially in the UD regime,
and calculations give a small size of this effect
\cite{Zimmermann91}. We propose that this is a signature of a new
collective mode emerging (or sharpening) below \Tc. The nature of
this mode is not clear at the moment. Its energy scale (100 - 270
meV) is much higher than $2\Delta$ ($\sim$ 20-30 meV
\cite{Nakano98}). The single-layer structure seems to exclude the
transverse plasmon scenario put forward for YBCO
\cite{Munzar99,Grueninger00}. However there is a similarity
between the effect we observe in LSCO and the increase of
conductivity at the same energies in YBCO
\cite{Homes95,Munzar99,Grueninger00}, so that the possibility that
the phenomena in these two compounds are common in origin cannot
be ruled out, even though the value of $\sigma_{1}$ as well as its
absolute change below \Tc\ is much larger in YBCO.

The occurrence of several collective modes has been recently
predicted by Lee and Nagaosa \cite{Lee02} from a mean-field
treatment of the $t$-$J$ model in the SU(2) formulation. One of
them, a so-called $\phi$ gauge mode, acquires a weak {\bf c}-axis
spectral weight in the low-temperature orthorhombic structure due
to the coupling to the buckling phonon mode. The energy scale of
this mode is the in-plane exchange $J$ ($\sim$ 100-150 meV), which
agrees well with our observation. Note, that this interpretation
does not involve the interplane charge transfer; the \bc-axis
infrared activity is due to the tilting of the oxygen octahedra.

The sum rule analysis can be applied to $\Delta_{s}\sigma_{1}$ in
the same manner as to $\Delta\sigma_{1}$. However, it directly
relates to the usual FGT sum rule only close to \Tc\ but not
automatically for $T\ll T_{c}$. From $D(T)$ (see
Fig.\ref{Fig2}(c)) we obtain $\Delta_{s}D$ of 27 and 130
$\Omega^{-1}$cm$^{-2}$K$^{-1}$ for $x$ = 0.12 and 0.15
respectively, denoted by the hatched area in Fig.\ref{Fig3}(a).

\begin{figure}
    \centerline{\psfig{figure=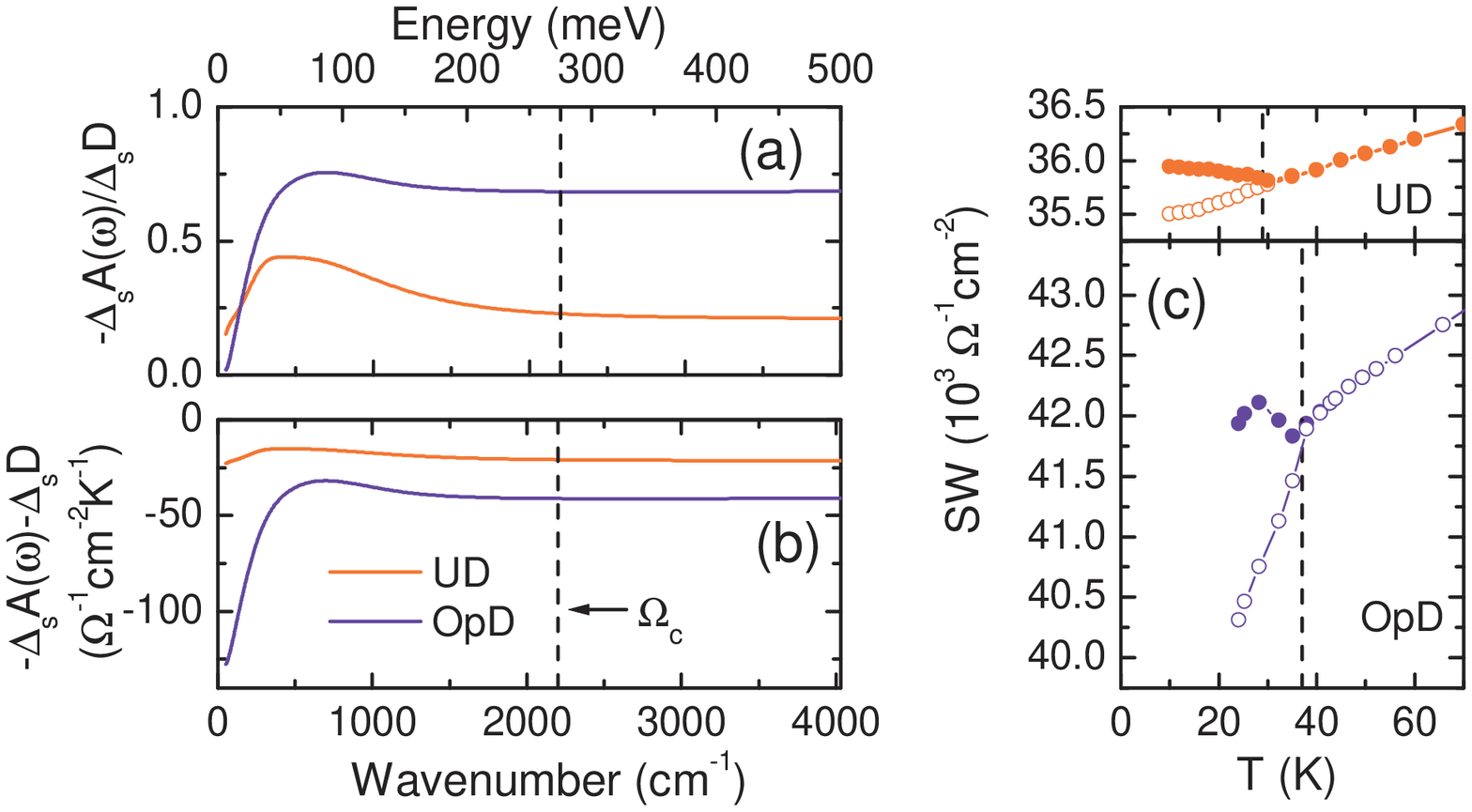,width=8.5cm,clip=}}
    \caption{The demonstration of the FGT sum rule violation in LSCO:
    (a) the relative finite-frequency SW $-\Delta_{s}A(\omega)/\Delta_{s}D$; (b)
    absolute SW $-\Delta_{s}A(\omega)- \Delta_{s}D$; (c) the temperature dependence of
    $A(\Omega_{c})$ (open circles) and $A(\Omega_{c})+D$ (solid
    circles). \Tc\ is marked by the vertical line.
    }
     \label{Fig4}
\end{figure}

Fig.\ref{Fig4}(a) shows the relative SW
$-\Delta_{s}A(\omega)/\Delta_{s}D$. It saturates above
$\Omega_{c}$ at the level of $\sim$ 0.2 for $x$=0.12 and $\sim$
0.7 for $x$=0.15, thus clearly showing the presence of the FGT sum
rule violation for both doping levels with relative values of
almost 80\% for the UD sample and 30\% for the OpD sample. This is
in excellent agreement with the tendency found in YBCO
\cite{Basov01}. In contrast to the relative violation, which is
stronger in the UD sample, the absolute decrease of the spectral
weight $A(\Omega_{c})+D$ is about 2 times larger for $x$=0.15
(Fig.\ref{Fig4}(b)). This can be also directly deduced from
Fig.\ref{Fig3}(c) using formula (\ref{DeltaEps}). Thus the
\bc-axis KE lowering is likely to be larger for the OpD sample,
which agrees with the prediction of Ref. \cite{Hirsch00}. The
positive $\Delta_{s}\sigma_{1}$ at higher frequencies
significantly increases the violation, especially in the UD
sample. One can even notice that the ratio between the positive
part of SW above the gap for two doping levels ($\sim$ 2) is close
to the ratio of $\Delta_{s}[A(\Omega_{c})+D]$. Finally we plot the
temperature dependence of $A(\Omega_{c})$ and $A(\Omega_{c})+D$
(Fig.\ref{Fig4}(b)). In the normal state the SW decreases with
cooling down, which has to be taken into account for a proper
formulation of the FGT sum rule. Below \Tc\ $A(\Omega_{c})$ starts
to decrease faster as a result of the gap opening, which is
partially screened by the intermediate-energy increase of SW. The
growth of $A(\Omega_{c})+D$ in the SC state is a signature of the
FGT sum rule violation.

In conclusion, our study, based on the combination of reflection
and transmission spectra, reveals a detailed picture of the
changes to the spectral weight in LSCO below \Tc\ for energies up
to 1.8 eV. The transition to the superconducting state is
accompanied not only by the gap opening at low frequencies
($2\Delta\sim$ 20-30 meV) but also by the increase of
$\sigma_{1}(\omega)$ at higher energies up to 270 meV. The latter
is interpreted as a possible occurrence of a new collective mode.
Its origin is not yet established. However the model by Lee and
Nagaosa predicts collective modes at similar energy. The total
low-frequency SW (including condensate) increases below \Tc; the
absolute value of this growth is higher in the OpD state, even
though the relative FGT sum rule violation is much larger for the
UD sample. Taking into account the new collective mode is decisive
for a quantitative understanding of the implications of the
sum-rule violation. Our results match qualitatively several
findings in YBCO, thus giving a hope that many of the '\bc-axis
effects' can be understood on the same basis in different cuprates
in spite of a diversity of their crystal structures. This
investigation was supported by the Netherlands Foundation for
Fundamental Research on Matter (FOM) with financial aid from the
Nederlandse Organisatie voor Wetenschappelijk Onderzoek (NWO) and
by DFG via SFB 608.


\begin{references}
\bibitem{TempCal}
Due to the limited thermal contact of the thin sample with the
holder the lowest sample temperature in the transmission
experiment was 10 K for $x$=0.12 and as high as 24 K for $x$=0.15.
It was calibrated by a comparison of the Josephson plasma edge in
the spectra of $R$ and $T$.
\bibitem{Spread}
The samples are slightly wedged, which suppresses the Fabry-Perot
fringes. To account for this effect we introduced the rectangular
thickness distribution with a width as an extra parameter. In this
approximation the interference pattern was reproduced quite well,
giving the thickness spreads of 0.5 \mkm\ for $x$=0.12 and 1.5
\mkm\ for $x$=0.15.
\bibitem{FormulaDeltaEps}
This formula follows from the KK relation $\Delta_{s}\epsilon_{1}(\omega)=
8\int_{0}^{\infty}\Delta_{s}\sigma_{1}(x)(x^2-\omega^2)^{-1}dx$
assuming $\Delta_{s}\sigma_{1}(x>\Omega_{c})=0$. Although the FGT
sum rule violation dictates that $\Delta_{s}\sigma_{1}(x)\neq 0$
at very high energies, the latter is very small up to at least 1.8
eV.
\end{references}
\end{document}